\documentclass[notitlepage]{revtex4-1}

\usepackage[utf8]{inputenc}
\usepackage[colorlinks=true,citecolor=blue,linkcolor=blue]{hyperref}
\usepackage[normalem]{ulem}
\usepackage{url}
\usepackage{graphicx,wrapfig,float,slashed,cancel}
\usepackage{amsmath,amssymb,epsfig,graphicx,xcolor,stmaryrd}
\usepackage{bm}
\usepackage{enumitem}
\usepackage{multirow}
\usepackage{soul}
\usepackage[utf8]{inputenc}
\usepackage{xcolor}
\usepackage{empheq}
\usepackage[many]{tcolorbox}

\definecolor{darkblue}{RGB}{1, 90, 173}

\begin{document}


\title{Interpreting the Newly Observed $\Xi(1720)$ State in the Spin-$\frac{3}{2}$ $\Xi$ Spectrum}

\author{K.~Azizi}
\email{ kazem.azizi@ut.ac.ir}
\thanks{Corresponding author}
\affiliation{Department of Physics, University of Tehran, North Karegar Avenue, Tehran
14395-547, Iran}
\affiliation{Department of Physics, Faculty of Engineering and Natural Sciences, Dogus University, Dudullu-\"{U}mraniye, 34775
Istanbul, T\"{u}rkiye}
\affiliation{School of Particles and Accelerators, Institute for Research in Fundamental Sciences (IPM) P.O. Box 19395-5531, Tehran, Iran}
\author{Y.~Sarac}
\email{yasemin.sarac@atilim.edu.tr}
\affiliation{Electrical and Electronics Engineering Department,
Atilim University, 06836 Ankara, T\"{u}rkiye}
\author{H.~Sundu}
\email{ hayriyesundu.pamuk@medeniyet.edu.tr}
\affiliation{Department of Physics Engineering, Istanbul Medeniyet University, 34700 Istanbul, T\"{u}rkiye}

\date{\today}

\preprint{}

\begin{abstract}

 We analyze the low lying spin-$\frac{3}{2}$ $\Xi$ spectrum by means of the two point QCD sum rule approach. The analysis is motivated by the BESIII observation of the $\Xi(1720)$ resonance, reported with mass $1721.0\pm5.2_{\mathrm{stat.}}\pm3.4_{\mathrm{syst.}}~\mathrm{MeV}$ and favored quantum numbers $J^P=\frac{3}{2}^{+}$. In this framework, we include the $1S$, $1P$, $2S$ and $2P$ configurations as possible low lying spin-$\frac{3}{2}$ cascade states. The extracted masses are $1527.53\pm111.38~\mathrm{MeV}$, $1615.30\pm50.98~\mathrm{MeV}$, $1727.52\pm42.39~\mathrm{MeV}$ and $1803.71\pm64.50~\mathrm{MeV}$ for the $1S$, $1P$, $2S$ and $2P$ states, respectively. The mass obtained for the $2S$ configuration agrees with the BESIII value within uncertainties and favors the assignment of $\Xi(1720)$ as the first radial spin-$\frac{3}{2}$ excitation of the $\Xi$ baryon.

\end{abstract}


\maketitle

\renewcommand{\thefootnote}{\#\arabic{footnote}}
\setcounter{footnote}{0}
\section{\label{sec:level1}Introduction}\label{intro}

The study of hadron spectra provides one of the useful ways to understand the dynamics of the strong interaction in the nonperturbative region. Although the quark model has been very successful in classifying hadrons, the spectrum of excited baryons is still not completely established. Among baryons, the $\Xi$ states are of particular interest because they carry strangeness $S=-2$ and their excited spectrum is much less known than those of the nucleon and $\Delta$ families. The available experimental information on excited $\Xi$ resonances is also rather limited. Several states predicted in this sector have not yet been observed, and the quantum numbers of some reported candidates are still uncertain. Therefore, each new observation in the cascade baryon sector is important for completing the baryon spectrum and for testing our understanding of nonperturbative QCD.

The spectrum of $\Xi$ baryons has been examined using a range of theoretical approaches, including nonrelativistic and relativistic quark models, Skyrme-type descriptions, chiral quark models, Regge phenomenology, QCD sumr rules, and QCD based methods \cite{Isgur1978,Oh2007,Lee2002,Pervin2008,Melde2008,Schat2002,Xiao2013,Chao1981,Capstick1986,CapstickRoberts2000,Glozman1998,GlozmanRiska1996,Menapara2021,Aliev:2016jnp}. These investigations have effectively described the established low lying cascade states, particularly $\Xi(1320)$ and $\Xi(1530)$. In contrast, the characterization of higher excitations remains ambiguous, as different models often yield varying mass predictions and quantum number assignments for the same resonance. This ongoing uncertainty demonstrates that the excited $\Xi$ spectrum is not yet fully understood, highlighting the need for further studies employing independent nonperturbative approaches.

The BESIII Collaboration recently reported the $\Xi(1720)$ state, adding another element to the still incomplete cascade baryon spectrum. In the partial wave analysis of the $J/\psi \to K^{-}\Sigma^{0}\bar{\Xi}^{+}+c.c.$ channel, this new doubly strange hyperon was identified as $\Xi(1720)$~\cite{BESIII:2026jcv}. The reported mass is $1721.0\pm5.2_{\mathrm{stat.}}\pm3.4_{\mathrm{syst.}}~\mathrm{MeV}$, while its width is $31.3\pm18.3_{\mathrm{stat.}}\pm15.4_{\mathrm{syst.}}~\mathrm{MeV}$. The spin-parity analysis favors the assignment $J^{P}=\frac{3}{2}^{+}$ for the observed state~\cite{BESIII:2026jcv}.

This newly observed state with its favored quantum numbers does not fit easily into the expectations of several conventional descriptions of the $\Xi$ spectrum, making it a challenging case for theoretical studies. In many model calculations, the first radial spin-$\frac{3}{2}^{+}$ $\Xi$ excitation is predicted considerably above the $\Xi(1720)$ mass region. For example, the relativistic quark-diquark model gives about $1966~\mathrm{MeV}$ for this state \cite{Faustov2015}, while the hypercentral constituent quark model with relativistic corrections predicts a value close to $1964~\mathrm{MeV}$ \cite{Menapara2021}. A similar result was obtained in the Regge phenomenology analysis \cite{Oudichhya2023}. Other theoretical studies also indicate that the corresponding positive parity excitation may appear in the higher mass region, roughly around $1.9$-$2.2~\mathrm{GeV}$ \cite{Loring:2001ky,Pervin2008,ChenMa2009,BGR2013}. Therefore, the experimentally observed $\Xi(1720)$, with favored quantum numbers $J^P=\frac{3}{2}^{+}$, lies below many theoretical expectations, providing a strong motivation for further analysis of this state.

To examine the possible nature of the $\Xi(1720)$ state from a QCD based perspective, we use the QCD sum rule framework, which is a widely applied nonperturbative tool in hadron phenomenology. In this approach, the correlation function is written in two representations: one in terms of hadronic degrees of freedom and the other in terms of quark and gluon degrees of freedom through the operator product expansion (OPE)~\cite{Shifman:1978bx,Shifman:1978y,Ioffe81}. After the Borel transformation is performed and the continuum contribution is subtracted, the mass and current coupling constant of the relevant state can be determined. In the present work, we study the spectroscopic parameters of spin-$\frac{3}{2}$ $\Xi$ baryons by including four possible configurations: the ground state, the first orbital excitation, the first radial excitation and the second orbital excitation, within the two-point QCD sum rule approach. Such an extension of the spectrum is quite natural in the strange baryon sector. Indeed, the PDG listings of the spin-$\frac{3}{2}$ $\Lambda$ states show that several resonances with the same spin but different parities appear in a relatively close mass region, including $\Lambda(1520)$, $\Lambda(1690)$ and $\Lambda(1890)$~\cite{ParticleDataGroup:2026aaa}. Based on this pattern, we calculate the masses and current coupling constants of four spin-$\frac{3}{2}$ $\Xi$ states with positive and negative parities and discuss whether the newly observed $\Xi(1720)$ state can be interpreted as a conventional $dss$ cascade baryon.

The rest of the paper is structured as follows. In Sec.~\ref{II}, we present the two-point correlation function and obtain the QCD sum rules for the spectroscopic parameters of the spin-$\frac{3}{2}$ $\Xi$ states. In Sec.~\ref{III}, the numerical results are given together with the working intervals of the auxiliary parameters, and our predictions are compared with available experimental information and theoretical results. Sec.~\ref{IV} contains the summary and concluding remarks.

\section{The QCD sum rule for the spin-$\frac{3}{2}$ $\Xi$ states}\label{II}

To investigate the $\Xi(1720)$ resonance, we start by assigning this state to the spin-$\frac{3}{2}$ $\Xi$ baryon family and calculating the corresponding spectroscopic parameters. The comparison of the obtained mass values with the BESIII measurement is expected to provide useful information about whether this state can be interpreted as a conventional cascade baryon. In the present analysis, we include four possible states corresponding to the $1S$, $1P$, $2S$ and $2P$ configurations of the spin-$\frac{3}{2}$ $dss$ system. We denote these states by $\Xi$, $\widetilde{\Xi}$, $\overline{\Xi}$ and $\widehat{\Xi}$, respectively. The states $\Xi$ and $\overline{\Xi}$ are considered as positive parity states, while $\widetilde{\Xi}$ and $\widehat{\Xi}$ represent the negative parity orbital excitations. To this end, we choose an interpolating current built from quark fields carrying the appropriate valence quark content and quantum numbers of the $\Xi$ baryons under consideration. The following interpolating current is used 
\begin{eqnarray}
J_{\mu}(x)=\frac{1}{\sqrt{3}}\epsilon^{ijk}
\left[
\left(s^{iT}(x)C\gamma_{\mu}s^{j}(x)\right)d^{k}(x)
+
\left(s^{iT}(x)C\gamma_{\mu}d^{j}(x)\right)s^{k}(x)
+
\left(d^{iT}(x)C\gamma_{\mu}s^{j}(x)\right)s^{k}(x)
\right],\label{Intcur}
\end{eqnarray}
inside the following two-point correlation function: 
\begin{eqnarray}
\Pi_{\mu\nu}(p)=i\int d^4x e^{ip\cdot x}\langle 0|T\{J_{\mu}(x)\bar{J}_{\nu}(0)\}|0\rangle.
\label{TwoPointCor}
\end{eqnarray}
In Eq.~(\ref{Intcur}), $i$, $j$ and $k$ denote color indices, while $C$ is the charge conjugation matrix. The fields $s(x)$ and $d(x)$ represent the strange and light quark fields, respectively, and $p$ in Eq.~(\ref{TwoPointCor}) is the four momentum carried by the baryon. Following the QCD sum rule procedure, the correlation function is analyzed in two forms. The hadronic representation is written in terms of the masses and current coupling constants of the relevant spin-$\frac{3}{2}$ $\Xi$ states, whereas the QCD representation is obtained from quark and gluon degrees of freedom through the operator product expansion. The required sum rules are then derived by matching these two representations after the Borel transformation and continuum subtraction.

On the hadronic side, the physical representation is constructed by saturating the correlation function with spin-$\frac{3}{2}$ states carrying the same quantum numbers as the interpolating current $J_{\mu}$. Since this current may couple to both positive and negative parity states, we include explicitly the ground state $(\Xi)$, the first orbital excitation $(\widetilde{\Xi})$, the first radial excitation $(\overline{\Xi})$ and the second orbital excitation $(\widehat{\Xi})$. After separating the pole contributions of these states, the physical side of the correlation function is written as
\begin{eqnarray}
	\Pi_{\mu\nu}^{\mathrm{Phys}}(p)&=&
	\frac{\langle 0|J_{\mu}|\Xi(p,s)\rangle
		\langle \Xi(p,s)|\bar{J}_{\nu}|0\rangle}
	{m^2-p^2}
	+
	\frac{\langle 0|J_{\mu}|\widetilde{\Xi}(p,s)\rangle
		\langle \widetilde{\Xi}(p,s)|\bar{J}_{\nu}|0\rangle}
	{\widetilde{m}^{2}-p^2}
	+
	\frac{\langle 0|J_{\mu}|\overline{\Xi}(p,s)\rangle
		\langle \overline{\Xi}(p,s)|\bar{J}_{\nu}|0\rangle}
	{\overline{m}^{2}-p^2}\nonumber\\
	&+&	
	\frac{\langle 0|J_{\mu}|\widehat{\Xi}(p,s)\rangle
		\langle \widehat{\Xi}(p,s)|\bar{J}_{\nu}|0\rangle}
	{\widehat{m}^{2}-p^2}
	+\cdots\,.
	\label{eq:masshadronicside1} 
\end{eqnarray}
The ellipsis denotes the contributions of higher resonances and continuum states. Eq.~(\ref{eq:masshadronicside1}) contains the pole terms of the four lowest spin-$\frac{3}{2}$ $\Xi$ configurations considered in this work, namely the ground state, the first orbital excitation, the first radial excitation and the second orbital excitation. For these four low lying states, we use the notation $\Xi$, $\widetilde{\Xi}$, $\overline{\Xi}$ and $\widehat{\Xi}$, and assign the masses $m$, $\widetilde{m}$, $\overline{m}$ and $\widehat{m}$ to them, respectively. In this notation, $\widetilde{\Xi}$ and $\widehat{\Xi}$ correspond to the negative parity orbital excitations, while $\Xi$ and $\overline{\Xi}$ represent the positive parity states. The relevant matrix elements are parametrized in terms of the current coupling constants and the Rarita Schwinger spinor $u_{\mu}(p,s)$ as
\begin{eqnarray}
	\langle 0|J_{\mu}|\Xi(p,s)\rangle
	&=& \lambda u_{\mu}(p,s),
	\nonumber\\
	\langle 0|J_{\mu}|\widetilde{\Xi}(p,s)\rangle
	&=& \widetilde{\lambda}\gamma_{5}u_{\mu}(p,s),
	\nonumber\\
	\langle 0|J_{\mu}|\overline{\Xi}(p,s)\rangle
	&=& \overline{\lambda}u_{\mu}(p,s),
	\nonumber\\
	\langle 0|J_{\mu}|\widehat{\Xi}(p,s)\rangle
	&=& \widehat{\lambda}\gamma_{5}u_{\mu}(p,s),
	\label{eq:matrixelements4}
\end{eqnarray}
where $\lambda$, $\widetilde{\lambda}$, $\overline{\lambda}$ and $\widehat{\lambda}$ denote the current coupling constants of the states $\Xi$, $\widetilde{\Xi}$, $\overline{\Xi}$ and $\widehat{\Xi}$, respectively.

After the definitions in Eq.~(\ref{eq:matrixelements4}) are inserted into the hadronic representation, the summation over the spin polarizations of the spin-$\frac{3}{2}$ baryons is performed. For a Rarita Schwinger spinor corresponding to the ground state $\Xi$, we use
\begin{eqnarray}
	\sum_s u_{\mu}(p,s)\bar{u}_{\nu}(p,s)
	&=&
	-\left(\slashed{p}+m\right)
	\left[
	g_{\mu\nu}
	-\frac{1}{3}\gamma_{\mu}\gamma_{\nu}
	-\frac{2p_{\mu}p_{\nu}}{3m^{2}}
	+\frac{p_{\mu}\gamma_{\nu}-p_{\nu}\gamma_{\mu}}{3m}
	\right].
	\label{eq:Rarita}
\end{eqnarray}
For the excited states, the same expression is used with the replacements $m\rightarrow\widetilde{m}$, $m\rightarrow\overline{m}$ and $m\rightarrow\widehat{m}$.

Although $J_{\mu}$ is chosen to interpolate spin-$\frac{3}{2}$ $\Xi$ states, it can also have nonzero overlap with spin-$\frac{1}{2}$ states. These unwanted contributions are parametrized as
\begin{eqnarray}
	\langle 0|J_{\mu}|\Xi_{{1/2}^{+}}(p)\rangle
	&=&
	A_{+}\left(\gamma_{\mu}-\frac{4p_{\mu}}{m_{{1/2}^{+}}}\right)u(p),
	\nonumber\\
	\langle 0|J_{\mu}|\Xi_{{1/2}^{-}}(p)\rangle
	&=&
	A_{-}\left(\gamma_{\mu}-\frac{4p_{\mu}}{m_{{1/2}^{-}}}\right)\gamma_{5}u(p),
	\label{eq:spinhalf}
\end{eqnarray}
with $A_{+}$ and $A_{-}$ being the coupling constants of the positive and negative parity spin-$\frac{1}{2}$ states. As follows from Eq.~(\ref{eq:spinhalf}), the unwanted spin-$\frac{1}{2}$ contributions appear in structures proportional to $\gamma_{\mu}$ and $p_{\mu}$. To avoid these contaminations, we arrange the Dirac matrices in a fixed order and focus on the structure $\slashed{p}g_{\mu\nu}$, which is free of spin-$\frac{1}{2}$ pollution. For this reason, in the numerical analysis we work with the coefficient of the $\slashed{p}g_{\mu\nu}$ structure. The hadronic side can then be written as
\begin{eqnarray}
	\Pi_{\mu\nu}^{\mathrm{Had}}(p)
	&=&
	\frac{\lambda^{2}}{p^{2}-m^{2}}
	\left(\slashed{p}+m\right)
	\left[
	g_{\mu\nu}
	-\frac{1}{3}\gamma_{\mu}\gamma_{\nu}
	-\frac{2p_{\mu}p_{\nu}}{3m^{2}}
	+\frac{p_{\mu}\gamma_{\nu}-p_{\nu}\gamma_{\mu}}{3m}
	\right]
	\nonumber\\
	&+&
	\frac{\widetilde{\lambda}^{2}}{p^{2}-\widetilde{m}^{2}}
	\left(\slashed{p}-\widetilde{m}\right)
	\left[
	g_{\mu\nu}
	-\frac{1}{3}\gamma_{\mu}\gamma_{\nu}
	-\frac{2p_{\mu}p_{\nu}}{3\widetilde{m}^{2}}
	-\frac{p_{\mu}\gamma_{\nu}-p_{\nu}\gamma_{\mu}}{3\widetilde{m}}
	\right]
	\nonumber\\
	&+&
	\frac{\overline{\lambda}^{2}}{p^{2}-\overline{m}^{2}}
	\left(\slashed{p}+\overline{m}\right)
	\left[
	g_{\mu\nu}
	-\frac{1}{3}\gamma_{\mu}\gamma_{\nu}
	-\frac{2p_{\mu}p_{\nu}}{3\overline{m}^{2}}
	+\frac{p_{\mu}\gamma_{\nu}-p_{\nu}\gamma_{\mu}}{3\overline{m}}
	\right]\nonumber\\
	&+&
	\frac{\widehat{\lambda}^{2}}{p^{2}-\widehat{m}^{2}}
	\left(\slashed{p}-\widehat{m}\right)
	\left[
	g_{\mu\nu}
	-\frac{1}{3}\gamma_{\mu}\gamma_{\nu}
	-\frac{2p_{\mu}p_{\nu}}{3\widehat{m}^{2}}
	-\frac{p_{\mu}\gamma_{\nu}-p_{\nu}\gamma_{\mu}}{3\widehat{m}}
	\right]
	+\cdots .
	\label{eq:PhyssSide}
\end{eqnarray}
The part proportional to $\slashed{p}g_{\mu\nu}$ is therefore obtained as
\begin{eqnarray}
	\Pi_{\mu\nu}^{\mathrm{Had}}(p)
	&=&
	\left[
	\frac{\lambda^{2}}{p^{2}-m^{2}}
	+
	\frac{\widetilde{\lambda}^{2}}{p^{2}-\widetilde{m}^{2}}
	+
	\frac{\overline{\lambda}^{2}}{p^{2}-\overline{m}^{2}}
	+
	\frac{\widehat{\lambda}^{2}}{p^{2}-\widehat{m}^{2}}
	\right]\slashed{p}g_{\mu\nu}
	+\cdots .
	\label{eq:CorFun1}
\end{eqnarray}
Performing the Borel transformation with respect to $-p^{2}$, we find
\begin{eqnarray}
	\widehat{\mathcal B}\Pi_{\mu\nu}^{\mathrm{Had}}(p)
	&=&
	\left[-
	\lambda^{2}e^{-m^{2}/M^{2}}
	-
	\widetilde{\lambda}^{2}e^{-\widetilde{m}^{2}/M^{2}}
	-
	\overline{\lambda}^{2}e^{-\overline{m}^{2}/M^{2}}
	-
	\widehat{\lambda}^{2}e^{-\widehat{m}^{2}/M^{2}}
	\right]\slashed{p}g_{\mu\nu}
	+\cdots .
	\label{eq:CorFunBorel}
\end{eqnarray}
The omitted terms represented by the ellipses in Eq.~(\ref{eq:CorFunBorel}) correspond to Lorentz structures other than $\slashed{p}g_{\mu\nu}$, which are not used in the present sum rule analysis, together with the contributions of higher resonances and continuum states.

For the QCD side of the correlation function, we employ the operator product expansion. In this part, the interpolating current in Eq.~(\ref{Intcur}) is substituted into the correlator given in Eq.~(\ref{TwoPointCor}). The resulting products of quark fields are then contracted by means of Wick's theorem. In this way, the correlation function is expressed through the light and strange quark propagators. For the spin-$\frac{3}{2}$ $\Xi$ current adopted in this work, the contracted expression takes the form
\begin{eqnarray}
	\Pi_{\mu\nu}^{\mathrm{QCD}}(p)
	&=&
	\frac{i}{3}\epsilon^{ijk}\epsilon^{i'j'k'}
	\int d^4x\, e^{ip\cdot x}
	\Bigg\{
	4S_s^{ki'}(x)\gamma_{\nu}\widetilde{S}_s^{ij'}(x)
	\gamma_{\mu}S_d^{jk'}(x)
	+4S_s^{kj'}(x)\gamma_{\nu}\widetilde{S}_d^{ji'}(x)
	\gamma_{\mu}S_s^{ik'}(x)
	\nonumber\\
	&
	+& 4S_d^{ki'}(x)\gamma_{\nu}\widetilde{S}_s^{ij'}(x)
	\gamma_{\mu}S_s^{jk'}(x)
	+2S_d^{kk'}(x)
	{\rm Tr}\left[
	S_s^{jj'}(x)\gamma_{\nu}\widetilde{S}_s^{ii'}(x)\gamma_{\mu}
	\right]
	\nonumber\\
	&
	-& 4S_s^{kk'}(x)
	{\rm Tr}\left[
	S_d^{ji'}(x)\gamma_{\nu}\widetilde{S}_s^{ij'}(x)\gamma_{\mu}
	\right]
	\Bigg\}.
	\label{eq:QCDside}
\end{eqnarray}
Here $S_d^{ij}(x)$ and $S_s^{ij}(x)$ denote the light and strange quark propagators, respectively, which have the following form
\begin{eqnarray}
	S_{q}^{ab}(x)&=&i\frac{x\!\!\!/}{2\pi^{2}x^{4}}\delta_{ab}-\frac{m_{q}}{4\pi^{2}x^{2}}\delta_{ab}-\frac{\langle
		\overline{q}q\rangle}{12}\Big(1-i\frac{m_{q}}{4}x\!\!\!/\Big)\delta_{ab}-\frac{x^{2}}{192}m_{0}^{2}\langle
	\overline{q}q\rangle\Big( 1-i\frac{m_{q}}{6}x\!\!\!/\Big)\delta_{ab}-\frac{ig_{s}G_{ab}^{\theta\eta}}{32\pi^{2}x^{2}}\Big[x\!\!\!/\sigma_{\theta\eta} +\sigma_{\theta\eta}x\!\!\!/ \Big]
	\nonumber\\&-&\frac{x\!\!\!/ x^{2}g_s^2}{7776}\langle
	\overline{q}q\rangle^2\delta_{ab}-\frac{x^4\langle
		\overline{q}q\rangle\langle
		g_s^2G^2\rangle}{27648}\delta_{ab}+\frac{m_q}{32\pi^2}[ln(\frac{-x^2\Lambda^2}{4})+2 \gamma_E]g_{s}G_{ab}^{\theta\eta}\sigma_{\theta\eta}+\cdots,\label{propagator}
\end{eqnarray}  
and
\begin{eqnarray}
	\widetilde{S}_{d(s)}^{ij}(x)
	&=&
	C\left[S_{d(s)}^{ij}(x)\right]^T C.
\end{eqnarray}
In the propagator, $\gamma_E \simeq 0.577$ is the Euler constant and $\Lambda$ is the QCD scale parameter.

After substituting these propagators into Eq.~(\ref{eq:QCDside}), the QCD side is decomposed in terms of the same Lorentz structures used in the hadronic representation as
\begin{eqnarray}
	\Pi_{\mu\nu}^{\mathrm{QCD}}(p)
	&=&
	\Pi^{\mathrm{QCD}}(p^{2})\slashed{p}g_{\mu\nu}
	+\cdots ,
	\label{eq:QCDStructures}
\end{eqnarray}
where $\cdots$ denotes the contributions coming from the remaining Lorentz structures. The invariant amplitude $\Pi^{\mathrm{QCD}}(p^{2})$ is then represented through the corresponding spectral density as
\begin{eqnarray}
	\Pi^{\mathrm{QCD}}(p^{2})
	&=&
	\int_{(m_d+2m_s)^2}^{\infty}
	\frac{\rho^{\mathrm{QCD}}(s)}{s-p^{2}} ds
	+\cdots,
	\label{eq:DispersionQCD}
\end{eqnarray}
where $\rho^{\mathrm{QCD}}(s)=\frac{1}{\pi}{\rm Im}\Pi^{\mathrm{QCD}}(s)$. The explicit form of the spectral density $\rho^{\mathrm{QCD}}(s)$ is quite lengthy. For this reason, we do not present it explicitly here, but use the resulting invariant amplitude in the numerical analysis after the Borel transformation and continuum subtraction. The next step is to improve the convergence of the OPE and reduce the effects of higher states. For this purpose, we apply the Borel transformation to the QCD side and subtract the continuum contribution using the standard quark hadron duality ansatz. The resulting Borel transformed QCD expression for the selected invariant amplitude is
\begin{eqnarray}
	\widehat{\mathcal B}\Pi^{\mathrm{QCD}}(M^2,s_0)
	&=&
	\int_{(m_d+2m_s)^2}^{s_0}
	\rho^{\mathrm{QCD}}(s)e^{-\frac{s}{M^2}}ds.
	\label{eq:QCDborel}
\end{eqnarray}
Here $\widehat{\mathcal B}\Pi^{\mathrm{QCD}}$ denotes the Borel transformed QCD invariant amplitude, and $s_0$ is the threshold parameter separating the contribution of the considered states from the higher resonances and continuum. Equating this expression to the corresponding hadronic amplitude gives the QCD sum rules for the four spin-$\frac{3}{2}$ $\Xi$ states.

Having completed the calculations on the hadronic and QCD sides, we match the coefficients of the $\slashed{p}g_{\mu\nu}$ structure. This gives
\begin{eqnarray}
	-\lambda^{2}e^{-m^{2}/M^{2}}
	-\widetilde{\lambda}^{2}e^{-\widetilde{m}^{2}/M^{2}}
	-\overline{\lambda}^{2}e^{-\overline{m}^{2}/M^{2}}
	-\widehat{\lambda}^{2}e^{-\widehat{m}^{2}/M^{2}}
	&=&
	\widehat{\mathcal B}\Pi^{\mathrm{QCD}}(M^2,s_0).
	\label{eq:sumrule_pslash}
\end{eqnarray} 

In the next section, this result is applied to extract both the masses and current coupling constants of the considered states. In the numerical analysis, we follow a successive subtraction method. First, the sum rules are used in the single pole approximation, where only the lowest $1S$ state is kept explicitly and the remaining contributions are included in the continuum. Once the mass and current coupling constant of the $1S$ state are obtained, these values are used as input in the analysis of the next state. For the first orbital excitation, the previously determined $1S$ contribution is included as input, and the $1S+1P+\mathrm{continuum}$ scheme is applied. The same strategy is then repeated for the $2S$ and $2P$ states. Thus, the $2S$ state is analyzed by using the $1S$ and $1P$ parameters, while the $2P$ state is extracted by using the previously obtained $1S$, $1P$ and $2S$ parameters as input.

\section{Numerical analysis}\label{III}

The sum rules derived above are now used to extract the spectroscopic parameters of the considered spin-$\frac{3}{2}$ $\Xi$ states. We study a four state system consisting of the ground state $\Xi$, the first orbital excitation $\widetilde{\Xi}$, the first radial excitation $\overline{\Xi}$ and the second orbital excitation $\widehat{\Xi}$. For the numerical analysis, we use the input parameters given in Table~\ref{tab:Inputs}. 
\begin{table}[h!]
	\begin{tabular}{|c|c|}
		\hline\hline
		Parameters & Values \\ \hline\hline
		$m_{d}$                                    & $4.67^{+0.48}_{-0.17}~\mathrm{MeV}$ \cite{ParticleDataGroup:2026aaa}\\
		$m_{s}$                                     & $93.4^{+8.6}_{-3.4}~\mathrm{MeV}$ \cite{ParticleDataGroup:2026aaa}\\
		$\langle \bar{q}q \rangle (1\mbox{GeV})$    & $(-0.24\pm 0.01)^3$ $\mathrm{GeV}^3$ \cite{Belyaev:1982sa}  \\
		$\langle \bar{s}s \rangle $                 & $0.8\langle \bar{q}q \rangle$ \cite{Belyaev:1982sa} \\
		$m_{0}^2 $                                  & $(0.8\pm0.1)$ $\mathrm{GeV}^2$ \cite{Belyaev:1982sa}\\
		$\langle \overline{q}g_s\sigma Gq\rangle$   & $m_{0}^2\langle \bar{q}q \rangle$ \\
		$\langle \frac{\alpha_s}{\pi} G^2 \rangle $ & $(0.012\pm0.004)$ $~\mathrm{GeV}^4 $\cite{Belyaev:1982cd}\\
		\hline\hline
	\end{tabular}%
	\caption{Numerical values of the input parameters used in the analysis.}
	\label{tab:Inputs}
\end{table} 
The numerical results also depend on the auxiliary parameters $M^2$ and $s_0$, which appear after the Borel transformation and continuum subtraction. These parameters are fixed according to the usual prescriptions of the QCD sum rule method. Since they are auxiliary quantities, the physical quantities extracted from the sum rules should remain stable against their variations within the chosen working intervals. In determining these intervals, we demand that the OPE shows a good convergence and that the pole contribution is sufficiently separated from the continuum contribution. Accordingly, the lower bound of $M^2$ is obtained from the convergence condition, whereas the upper bound is constrained by the pole dominance requirement. The parameter $s_0$ is chosen by taking into account the expected position of the next higher resonance above the state under consideration at each step.

\begin{figure}[h!]
	\begin{center}
		\includegraphics[totalheight=5cm,width=7cm]{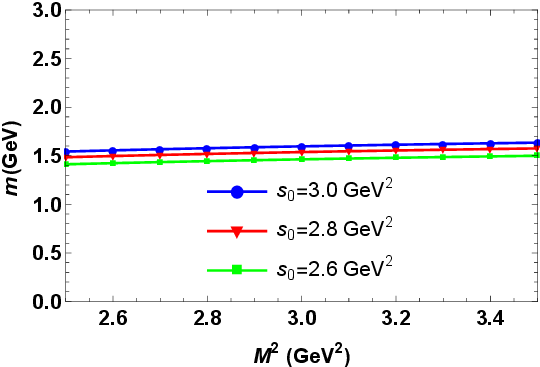}
		\includegraphics[totalheight=5cm,width=7cm]{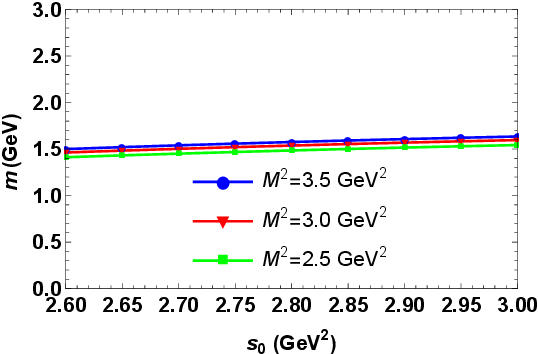}
	\end{center}
	\caption{\textbf{Left:} Variation of the ground state $\Xi$ mass with $M^2$ for different $s_0$ values.
		\textbf{Right:}  Variation of the ground state $\Xi$ mass with $s_0$ for different $M^2$ values.}
	\label{gr:MassMsqS01S}
\end{figure}

\begin{figure}[h!]
	\begin{center}
		\includegraphics[totalheight=5cm,width=7cm]{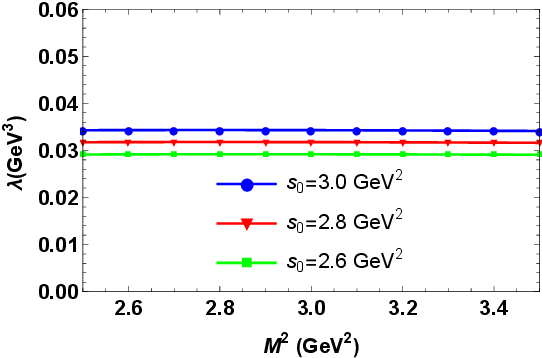}
		\includegraphics[totalheight=5cm,width=7cm]{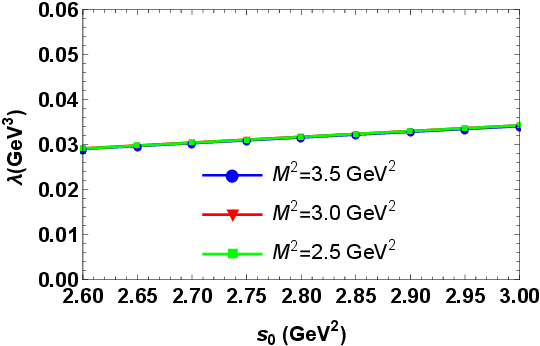}
	\end{center}
	\caption{\textbf{Left:} Variation of the ground state $\Xi$ current coupling constant with $M^2$ for different $s_0$ values.
		\textbf{Right:} Variation of the ground state $\Xi$ current coupling constant with $s_0$ for different $M^2$ values.}
	\label{gr:lamMsqS01S}
\end{figure}

\begin{figure}[h!]
	\begin{center}
		\includegraphics[totalheight=5cm,width=7cm]{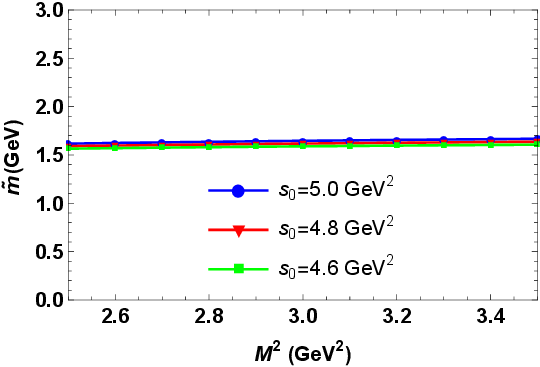}
		\includegraphics[totalheight=5cm,width=7cm]{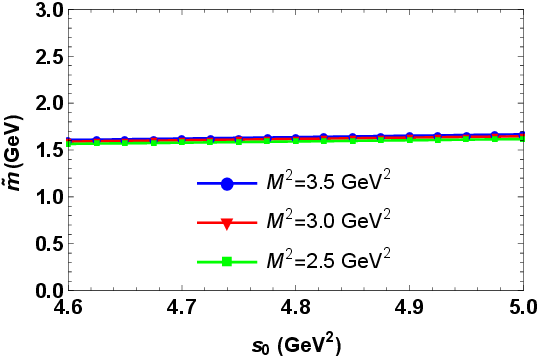}
	\end{center}
	\caption{\textbf{Left:} Variation of the mass of the first orbital excitation $\widetilde{\Xi}$ with $M^2$ for different $s_0$ values.
		\textbf{Right:} Variation of the mass of the first orbital excitation $\widetilde{\Xi}$ with $s_0$ for different $M^2$ values.}
	\label{gr:MassMsqS01P}
\end{figure}

\begin{figure}[h!]
	\begin{center}
		\includegraphics[totalheight=5cm,width=7cm]{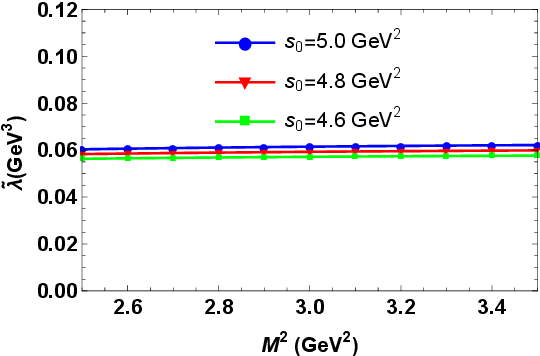}
		\includegraphics[totalheight=5cm,width=7cm]{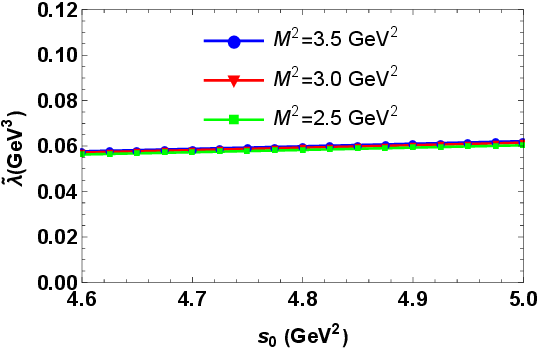}
	\end{center}
	\caption{\textbf{Left:} Variation of the current coupling constant of the first orbital excitation $\widetilde{\Xi}$ with $M^2$ for different $s_0$ values.
		\textbf{Right:} Variation of the current coupling constant of the first orbital excitation $\widetilde{\Xi}$ with $s_0$ for different $M^2$ values.}
	\label{gr:lamMsqS01P}
\end{figure}

\begin{figure}[h!]
	\begin{center}
		\includegraphics[totalheight=5cm,width=7cm]{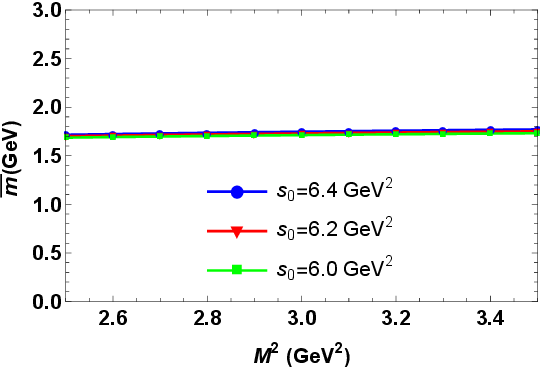}
		\includegraphics[totalheight=5cm,width=7cm]{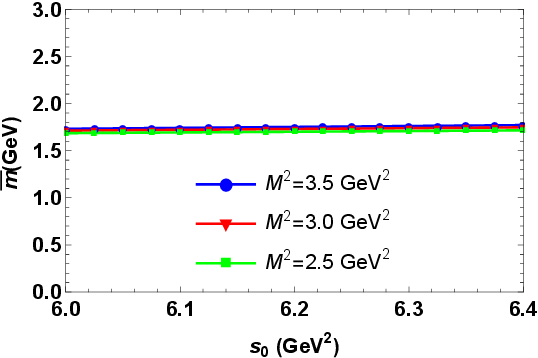}
	\end{center}
\caption{\textbf{Left:} Variation of the mass of the first radial excitation $\overline{\Xi}$ with $M^2$ for different $s_0$ values.
	\textbf{Right:} Variation of the mass of the first radial excitation $\overline{\Xi}$ with $s_0$ for different $M^2$ values.}
	\label{gr:MassMsqS02S}
\end{figure}

\begin{figure}[h!]
	\begin{center}
		\includegraphics[totalheight=5cm,width=7cm]{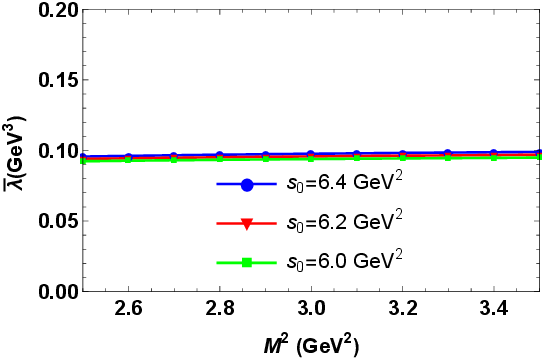}
		\includegraphics[totalheight=5cm,width=7cm]{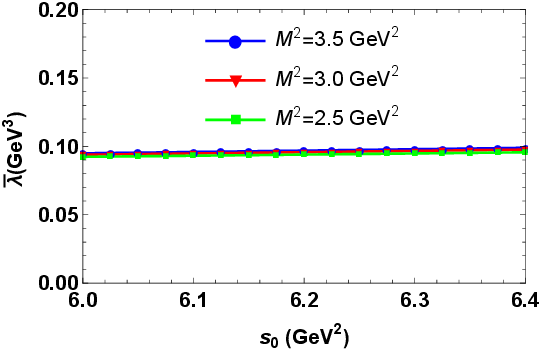}
	\end{center}
	\caption{\textbf{Left:} Variation of the current coupling constant of the first radial excitation $\overline{\Xi}$ with $M^2$ for different $s_0$ values.
		\textbf{Right:} Variation of the current coupling constant of the first radial excitation $\overline{\Xi}$ with $s_0$ for different $M^2$ values.}
	\label{gr:lamMsqS02S}
\end{figure}

\begin{figure}[h!]
	\begin{center}
		\includegraphics[totalheight=5cm,width=7cm]{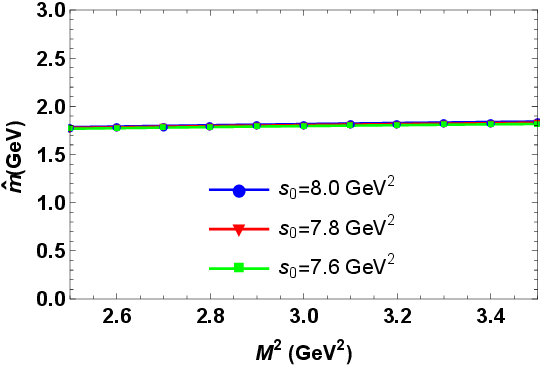}
		\includegraphics[totalheight=5cm,width=7cm]{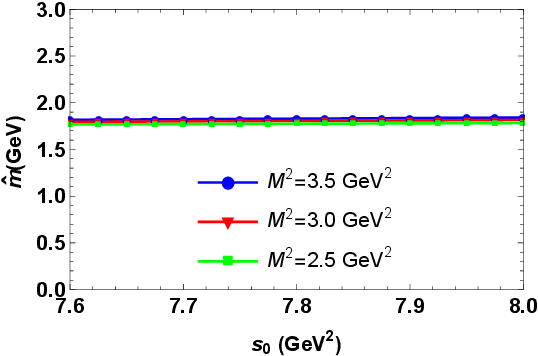}
	\end{center}
\caption{\textbf{Left:} Variation of the mass of the second orbital excitation $\widehat{\Xi}$ with $M^2$ for different $s_0$ values.
	\textbf{Right:} Variation of the mass of the second orbital excitation $\widehat{\Xi}$ with $s_0$ for different $M^2$ values.}
	\label{gr:MassMsqS02P}
\end{figure}

\begin{figure}[h!]
	\begin{center}
		\includegraphics[totalheight=5cm,width=7cm]{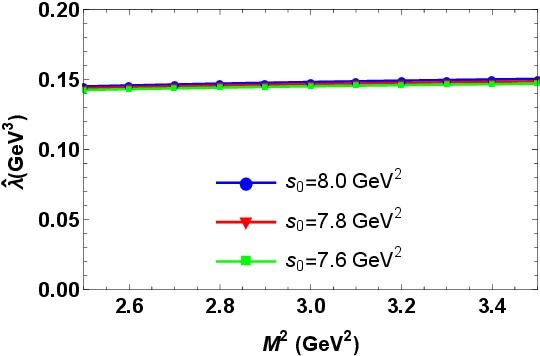}
		\includegraphics[totalheight=5cm,width=7cm]{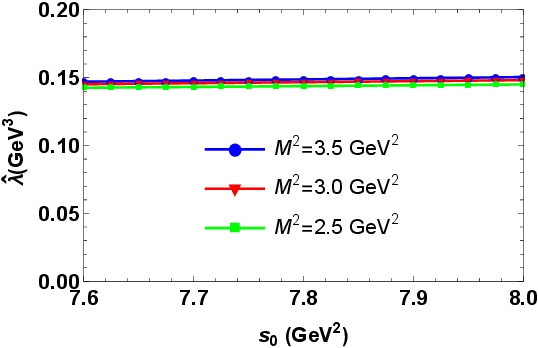}
	\end{center}
\caption{\textbf{Left:} Variation of the current coupling constant of the second orbital excitation $\widehat{\Xi}$ with $M^2$ for different $s_0$ values.
	\textbf{Right:} Variation of the current coupling constant of the second orbital excitation $\widehat{\Xi}$ with $s_0$ for different $M^2$ values.}
	\label{gr:lamMsqS02P}
\end{figure}

Following the criteria mentioned above, we determine the working regions of the Borel mass parameter and the continuum threshold for each step of the analysis. These intervals, as well as the extracted spectroscopic parameters, are shown in Table~\ref{tab:results_pslash}.
\begin{table*}[h!]
	\centering
	\begin{tabular}{|c|c|c|c|c|c|c|}
		\hline
		State & Assignment & $s_0~(\mathrm{GeV}^2)$ & $M^2~(\mathrm{GeV}^2)$ & PC $(\%)$ & Mass $(\mathrm{MeV})$ & Current coupling $(\mathrm{GeV}^3)$ \\
		\hline \hline
		$\Xi$ & $1S$ & $2.6-3.0$ & $2.5-3.5$ & $8.4-18$ & $1527.53\pm111.38$ & $0.0317\pm0.0025$ \\
		\hline
		$\widetilde{\Xi}$ & $1P$ & $4.6-5.0$ & $2.5-3.5$ & $20-37$ & $1615.30\pm50.98$ & $0.0592\pm0.0029$ \\
		\hline
		$\overline{\Xi}$ & $2S$ & $6.0-6.4$ & $2.5-3.5$ & $30-51$ & $1727.52\pm42.39$ & $0.0956\pm0.0033$ \\
		\hline
		$\widehat{\Xi}$ & $2P$ & $7.6-8.0$ & $2.5-3.5$ & $41-64$ & $1803.71\pm64.50$ & $0.1464\pm0.0039$ \\
		\hline
	\end{tabular}
	\caption{Working intervals of the auxiliary parameters, pole contribution values and extracted spectroscopic parameters of the spin-$\frac{3}{2}$ $\Xi$ states obtained from the $\slashed{p}g_{\mu\nu}$ structure.}
	\label{tab:results_pslash}
\end{table*}
The uncertainties presented in Table~\ref{tab:results_pslash} originate from the input parameters entering the QCD side and from the determination of the working intervals for $M^2$ and $s_0$. Since the analysis is performed step by step, different continuum thresholds are used for the $1S$, $1P$, $2S$ and $2P$ states. For the ground state, the physical side is approximated by the $1S+\mathrm{continuum}$ scheme. For the first orbital excitation, the previously determined $1S$ contribution is used as input, and the $1S+1P+\mathrm{continuum}$ scheme is applied. The $2S$ and $2P$ states are obtained in the same way by using the lower state parameters determined in the previous steps. To check the dominance of the explicitly included low lying contributions over the higher states and continuum, we use the pole contribution ratio
\begin{eqnarray}
	\mathrm{PC}(M^2,s_0)
	&=&
	\frac{\widehat{\mathcal B}\Pi^{\mathrm{QCD}}(M^2,s_0)}
	{\widehat{\mathcal B}\Pi^{\mathrm{QCD}}(M^2,\infty)} .
	\label{eq:PC}
\end{eqnarray}
A sufficiently large value of this quantity indicates that the selected pole contribution is dominant in the chosen working interval. The pole contribution values corresponding to these working intervals obtained in the selected working regions of $M^2$ and $s_0$ are also included in Table~\ref{tab:results_pslash}. The PC values listed in Table~\ref{tab:results_pslash} indicate that the explicitly included low lying pole contributions are sufficiently separated from the higher resonances and continuum. Therefore, the continuum subtracted sum rules can be reliably used to determine the masses and current coupling constants of the considered states. The dependence of the masses and current coupling constants on the Borel parameter and continuum threshold is displayed in Figs.~\ref{gr:MassMsqS01S}-\ref{gr:lamMsqS02P}. 
As is seen from Table~\ref{tab:results_pslash}, the obtained masses form a natural low lying spin-$\frac{3}{2}$ $\Xi$ spectrum. The $1S$ state appears around the $\Xi(1530)$ region, while the $2S$ state lies close to the newly observed $\Xi(1720)$ resonance. The predicted $1P$ and $2P$ masses also fall in the regions where negative parity cascade excitations may appear.

To better understand the physical interpretation of the obtained masses, we compare our results with previous theoretical predictions for spin-$\frac{3}{2}$ $\Xi$ baryons. The comparison is presented in Table~\ref{tab:comparison}. In the column of Ref.~\cite{Menapara2021}, the two values denote the masses obtained without and with the first order correction, respectively. In the column of Ref.~\cite{Melde2008}, the two values correspond to the GBE and OGE models. The comparison is made according to the spin-parity assignment and the corresponding low lying mass ordering, rather than a strict one to one identification of the internal excitation labels used in different models.
\begin{table*}[h!]
	\centering
	\tiny
	\renewcommand{\arraystretch}{1.15}
	\setlength{\tabcolsep}{1.6pt}
	\resizebox{\textwidth}{!}{%
		\begin{tabular}{|c|c|c|c|c|c|c|c|c|c|c|c|c|c|}
			\hline
			State
			& Assignment
			& This work
			& \cite{Oh2007}
			& \cite{Pervin2008}
			& \cite{Melde2008}
			& \cite{Capstick1986}
			& \cite{Menapara2021}
			& \cite{Faustov2015}
			& \cite{ChenMa2009}
			& \cite{BGR2013}
			& \cite{Loring:2001ky}
			& \cite{Bijker2000}
			& \cite{Santopinto2015} \tabularnewline
			\hline
			
			$\Xi$
			& $1S,\frac{3}{2}^{+}$
			& $1527.53\pm111.38$
			& $1539$
			& $1520$
			& $1528/1516$
			& $1505$
			& $1531/1524$
			& $1518$
			& $1526$
			& $1553\pm18$
			& $1539$
			& $1524$
			& $1552$ \tabularnewline
			\hline
			
			$\widetilde{\Xi}$
			& $1P,\frac{3}{2}^{-}$
			& $1615.30\pm50.98$
			& $1820$
			& $1759$
			& $1792/1894$
			& $1785$
			& $1871/1873$
			& $1764$
			& $1801$
			& $1906\pm29$
			& $1780$
			& $1828$
			& $1861$ \tabularnewline
			\hline
			
			$\overline{\Xi}$
			& $2S,\frac{3}{2}^{+}$
			& $1727.52\pm42.39$
			& $2120$
			& $1934$
			& $ $
			& $2045$
			& $1971/1964$
			& $1966$
			& $1952$
			& $2228\pm40$
			& $1988$
			& $1878$
			& $1653$ \tabularnewline
			\hline
			
			$\widehat{\Xi}$
			& $2P,\frac{3}{2}^{-}$
			& $1803.71\pm64.50$
			& $ $
			& $1826$
			& $ $
			& $ $
			& $1879/1881$
			& $1798$
			& $1918$
			& $1894\pm38$
			& $1873$
			& $1869$
			& $1971$ \tabularnewline
			\hline
			
		\end{tabular}%
	}
	\normalsize
	\caption{Comparison of our mass predictions with theoretical results for the spin-$\frac{3}{2}$ $\Xi$ states. All masses are given in MeV.}
	\label{tab:comparison}
\end{table*}
The comparison in Table~\ref{tab:comparison} shows that our ground state result is consistent with the theoretical predictions for the lowest spin-$\frac{3}{2}^{+}$ $\Xi$ baryon. It is also compatible with the established $\Xi(1530)$ state, whose mass is given as $1531.80\pm0.32~\mathrm{MeV}$ by the PDG \cite{ParticleDataGroup:2026aaa}. For the first orbital excitation, our prediction is lower than most of the quoted theoretical estimates; however, it is close to the $\Xi(1620)$ mass region and to the Belle result $1610.4\pm6.0^{+5.9}_{-3.5}~\mathrm{MeV}$ \cite{Belle2019Xi1620}. Since the spin-parity assignment of $\Xi(1620)$ is not yet definite, this result may be interpreted as a possible indication of a lower lying $J^P=\frac{3}{2}^{-}$ cascade state. For the $2S$ state, our result differs from most theoretical predictions, which generally place the first radial spin-$\frac{3}{2}^{+}$ $\Xi$ excitation at higher masses. Instead, our value agrees well with the newly observed $\Xi(1720)$ resonance reported by BESIII with mass $1721.0\pm5.2\pm3.4~\mathrm{MeV}$ and favored quantum numbers $J^P=\frac{3}{2}^{+}$ \cite{BESIII:2026jcv}. This agreement supports the interpretation of $\Xi(1720)$ as the first radial excitation of the spin-$\frac{3}{2}$ $\Xi$ baryon. Finally, the predicted $2P$ mass is close to the $\Xi(1820)$ state, which is listed by the PDG with $J^P=\frac{3}{2}^{-}$ and mass $1823\pm5~\mathrm{MeV}$ \cite{ParticleDataGroup:2026aaa}.

\section{Summary and conclusion}\label{IV}

In the present study, motivated by the newly observed $\Xi(1720)$ state, we analyzed the low lying spin-$\frac{3}{2}$ $\Xi$ spectrum within the two-point QCD sum rule method. For this purpose, we constructed the correlation function using a spin-$\frac{3}{2}$ interpolating current with the quark content $dss$. In the hadronic representation, four states were included explicitly: the ground state $\Xi(1S)$, the first orbital excitation $\widetilde{\Xi}(1P)$, the first radial excitation $\overline{\Xi}(2S)$ and the second orbital excitation $\widehat{\Xi}(2P)$. The sum rules were obtained by matching the coefficients of the $\slashed{p}g_{\mu\nu}$ structure from the hadronic and QCD sides after applying the Borel transformation and continuum subtraction.

Our numerical results give $1527.53\pm111.38~\mathrm{MeV}$, $1615.30\pm50.98~\mathrm{MeV}$, $1727.52\pm42.39~\mathrm{MeV}$ and $1803.71\pm64.50~\mathrm{MeV}$ for the $1S$, $1P$, $2S$ and $2P$ spin-$\frac{3}{2}$ $\Xi$ states, respectively. The obtained ground state mass is compatible with the $\Xi(1530)$ resonance, while the $2P$ prediction lies close to the $\Xi(1820)$ region. The central result of the present work is the mass obtained for the $2S$ state. The BESIII Collaboration reported the newly observed $\Xi(1720)$ resonance with $m_{\Xi(1720)}=1721.0\pm5.2_{\mathrm{stat.}}\pm3.4_{\mathrm{syst.}}~\mathrm{MeV}$  and favored quantum numbers $J^P=\frac{3}{2}^{+}$ \cite{BESIII:2026jcv}. Since our prediction for the $2S$ mass, $\overline{m}=1727.52\pm42.39~\mathrm{MeV}$, is consistent with this experimental value within uncertainties, the present analysis supports the interpretation of $\Xi(1720)$ as the first radial excitation of the spin-$\frac{3}{2}$ $\Xi$ baryon.

The comparison performed in Sec.~\ref{III} shows that our results differ from some quark-model predictions, especially for the radial excitation. While several previous studies place the first radial spin-$\frac{3}{2}^{+}$ $\Xi$ state near the $\Xi(1950)$ region, our result favors a lower mass compatible with the BESIII $\Xi(1720)$ observation. In addition, the predicted $1P$ mass suggests that a negative parity partner may appear near the $\Xi(1620)$ region. Therefore, the present results may be useful for future experimental analyses aiming to clarify the spin-parity assignments of excited cascade baryons and for further theoretical studies of strange baryon spectroscopy.


\section*{ACKNOWLEDGEMENTS}
 K. Azizi is grateful to Iran National Science Foundation (INSF) for the  partial financial support provided under the elites Grant No.40405095.

\section*{Declaration of AI-assisted language editing}

The authors used ChatGPT, developed by OpenAI, only for language editing and improving the readability of the manuscript. The authors reviewed and edited the manuscript and take full responsibility for its scientific content.


\bibliographystyle{apsrev4-1}  
\bibliography{Xi_1720biblio}


\end{document}